# A Comparison of Experimental Measurements and Computational Predictions of a Deep-V Planing Hull

T.C. Fu[1], T. Ratcliffe[1], T.T. O'Shea[2], K.A. Brucker[2], R.S. Graham[2], D.C. Wyatt[2] and D.G. Dommermuth[2]

([1]Carderock Division, NSWC, USA; [2]SAIC, La Jolla, CA, USA)

**ABSTRACT**

In order to support development of computational fluid dynamics codes for high-speed, small-craft applications, laboratory experiments were performed on a representative of Deep-V planing craft model. The measurements included resistance, sinkage and trim, hull pressure measurements, longitudinal wave-cuts, and bow-wave and stern-wake topologies. The model was towed in calm water over a speed range of 1.78 to 14.2 m/s (5.8 to 46.6 ft/s) corresponding to a Froude number range of 0.31 to 2.5. At planing speeds, +8 m/s (+26.2 ft/s), the model was run with the addition of trim tabs set at two different angles, 7 and 13 degrees. Photographic documentation, including still photographs and video, was recorded during the collection of all the data. Numerical simulation of the flow field was performed utilizing the Numerical Flow Analysis (NFA) code and compared to the model test results.

## INTRODUCTION

The qualitative and quantitative characterization of the complex multiphase free-surface flow field generated by a Deep-V monohull planing boat at high Froude numbers is a challenge to both measure experimentally and simulate numerically. Free-Surface Computational Fluid Dynamics (CFD) codes have been developed, primarily to predict the flow around displacement hull ships. In order to extend the use of these codes for predicting the flow around planing craft, a test program was developed to obtain resistance, sinkage and trim, hull pressures, and free-surface topologies on a representative Deep-V planing craft. This data can be used in the development and validation of flow codes for this type of hull form.

Due to the complexity of planing craft hydrodynamics, it was desirable to maximize the model size (thus minimizing scaling errors) while still being able to obtain a wide Froude number range (0.31 to 2.5) in a tow tank. The model was tested on Carriage 5 at the David Taylor Model Basin, Naval Surface Warfare Center Carderock Division, which has a top-end test speed of 50 knots. The basin dimensions are: length of 904 m (2966 ft) and width of 6.4 m (21 ft). The basin has a shallow section with a depth of 3 m (10 ft) and a deep section with a depth of 4.9 m (16 ft). The majority of the experimental measurements were performed in the deep section of the tow tank. The model was tested in both calm water and regular waves over a speed range of 1.78 to 14.2 m/s (5.8 to 46.6 ft/s) corresponding to a Froude number range of 0.31 to 2.5. Only the calm water data will be discussed in this paper. At planing speeds (+8 m/s), the model was run with the addition of trim tabs at tab angles, of 7 and 13 degrees.

## EXPERIMENT

While the planing craft model test program that was undertaken focused on collecting a wide range of types of measurements for comparison with CFD predictions, the comparison below will focus on (1) resistance, sinkage and trim, (2) wave topology, and (3) flow visualization.

**Model**
The Deep-V planing hull chosen for this test program was representative of a monohull planing hull craft. The model size was chosen to be as large as practical for testing on Carriage 5 at NSWCCD to minimize scale effects, while covering as wide a speed range as possible, up to 14 m/s (46.6 ft/s). A Deep-V monohull planing craft model was built of pine lifts in the model shop at the Naval Surface Warfare

Center, Carderock Division (NSWCCD). The model was painted yellow and station lines were added to the hull bottom and sides. A checkerboard grid comprised of one inch squares was applied to the hull bottom, and waterline marks were added to the station lines on the hull sides as visual aids for calculation of wetted surface area. Figure 1 shows photographs of the model tested.

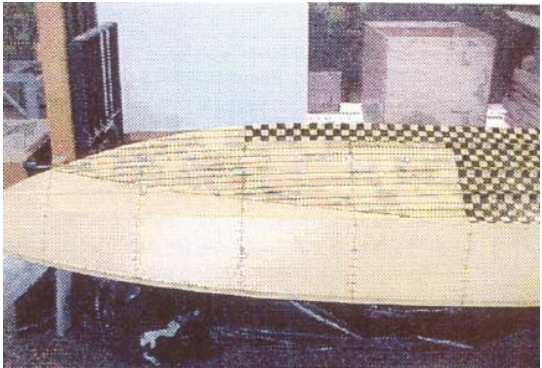

a)

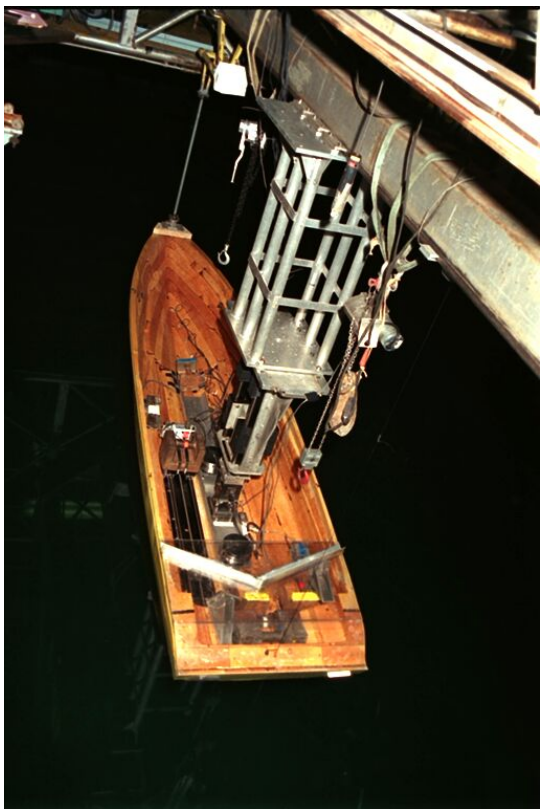

b)

**Figure 1**: a) The model with the station lines and grid markings and b) mounted on the towing carriage.

In order to collect measurements in both calm water and in waves, both static and dynamic ballasting of the model was performed. Additionally, in contrast to displacement vessels with fixed propulsors and rudders, planing craft are typically propelled by steerable propulsors – outboard motors, stern drives/outdrives, waterjets, etc. The thrust provided by these propulsors has horizontal and vertical components, and the vertical component becomes significant as the craft trims. Additionally, the thrust line can be at or even below the keel line. Both the horizontal and vertical thrust forces anticipated on a full-scale craft must be accurately represented at model scale. However, only horizontal force is applied to the model. To simulate the vertical component of the thrust, an upward force, an amount of weight equivalent in magnitude to the anticipated vertical force component is removed from the model without shifting the longitudinal center of gravity (LCG). To accurately represent application of propulsor thrust to a full-scale craft, the model should be towed at the point at which thrust is applied - the intersection of the LCG and the thrust (shaft) line. In most cases, a planing model cannot be towed at this location, since the thrust line is too low. The model is still towed at the LCG, but at a higher location, creating a moment which does not exist at full scale. This artificial moment was compensated while ballasting the model.

Model testing was divided into two test periods. The first period focused on the resistance and sinkage and trim measurements; the focus of the second test period was the free-surface wave field measurement. Resistance and sinkage and trim were also measured during the second test period to insure reproducible results. Tables 1 and 2 show the test conditions performed in each test period.

Additionally, wetted-surface calculations were made for each test speed. Wetted-bottom area was determined using underwater photos and the checkerboard grid (see Figure 2a). Wetted side area was determined using side photos (Figure 2b); the area was approximated as a triangle.

| Table 1: Test Matrix - First Testing Period |||| 
|---|---|---|---|
| Test number | Sea State | Description | Data Collected |
| 1 | Calm | Check out test | |
| 2 | Calm | 8.9, 11.8 m/s, 7° tabs | RST |
| 3 | Calm | 8.9 m/s, 13° tabs | RST |
| 4 | Calm | 3, 8.9, 11.8 m/s, 13° tabs | RST |
| 5 | Calm | 1.8, 2.4, 3 m/s, No tabs | RST |
| 6 | Calm | Trim checks | |
| 7 | Calm | 8.9, 11.8 m/s, 13° tabs | RST |
| 8 | Calm | 8.9, 11.8 m/s, 13° tabs | RST |
| 9 | Calm | 8.9, 11.8 m/s, 7° tabs | RST |
| 10 | Calm | 3, 14.2 m/s 13° tabs | RST |

**Notes:** RST: resistance, sinkage and trim, Tests #11-14 were in waves tests.

| Table 2: Test Matrix - Second Test Period |||| 
|---|---|---|---|
| Test number | Sea State | Description | Data Collected |
| 15 | Calm | 1.8 m/s, No tabs | RST, ST, QVIZ |
| 16 | Calm | 1.8, 3 m/s, No tabs | RST, ST, QVIZ |
| 17 | Calm | 1.8, 3 m/s, No tabs | RST, ST, QVIZ |
| 18 | Calm | 8.9 m/s, 13° tabs | RST, ST, QVIZ |
| 19 | Calm | 8.9 m/s, 7° tabs | RST, ST, QVIZ |
| 20 | Calm | 8.9 m/s, 7° tabs | RST, ST, QVIZ |
| 21 | Calm | 8.9 m/s, 13° tabs | RST, ST, QVIZ |
| 22 | Calm | 11.8 m/s, 13° tabs | RST, ST, QVIZ |
| 23 | Calm | 8.9 m/s, No tabs | RST, ST, QVIZ |
| 24 | Calm | 8.9 m/s, No tabs | RST, ST, QVIZ |
| 25 | Calm | 11.8 m/s, 13° tabs | RST, ST, QVIZ |
| 26 | Calm | 2.4 m/s, No tabs | RST, ST, QVIZ |
| 27 | Calm | 1.8, 2.4, 3 m/s, No tabs | RST, ST, QVIZ |

**Notes:** RST: resistance, sinkage and trim; ST: stern topography; QVIZ: quantitative visualization

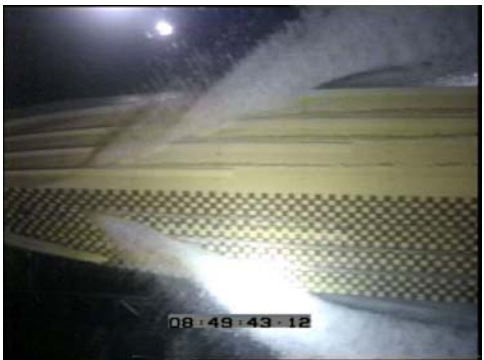

a) Bottom view

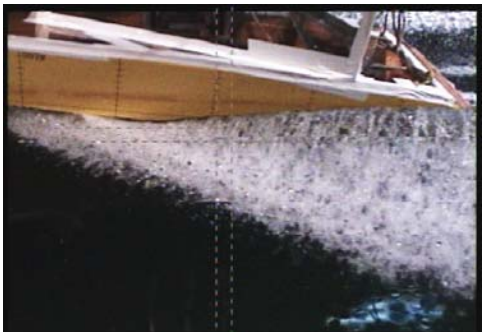

b) Side view

**Figure 2**: Sample images used to compute wetted surface area: a) the bottom view and b) the side view at a speed of 8.9 m/s (29.2 ft/s) and 7 degree trim tab angle.

## EXPERIMENTAL RESULTS

### Resistance, Trim & Heave

The averaged resistance (drag), trim angle, and heave at the tow post are shown in Figures 3-5. As can be seen there is enormous variation in values at the model speed of 2.96 m/s (9.7 ft/s). At this speed, the model is starting the transition from displacement to planing speed, and its position in the water is constantly changing. Therefore, the wide variation in data values is not unexpected. The addition of trim tabs increases drag at higher speeds (drag is higher for 13 degree tabs than 7 degree tabs), and lowers trim, as expected.

### Stern Topology

Figure 6 shows the stern topologies for four speeds of 1.8, 2.4, 3, and 9 m/s (5.8, 7.8, 9.7 and 29.2 ft/s) with no trim tabs. There is little variation transversely for the 1.8 and 2.4 m/s (5.8 and 7.8 ft/s) speeds. At 3 m/s (9.7 ft/s), the flow field varies some in Y/L at small X/L, but as X/L increases the topology becomes fairly homogeneous with respect to Y/L. At 9 m/s (29.2 ft/s), the topology has a significant gradation with respect to Y/L. This gradation does not degrade with X/L like in the case of the 3 m/s (9.7 ft/s) speed. Figure 7 is a comparison of the stern topography for a speed of 9 m/s (29.2 ft/s) with the 7 degree and 13 degree tabs.

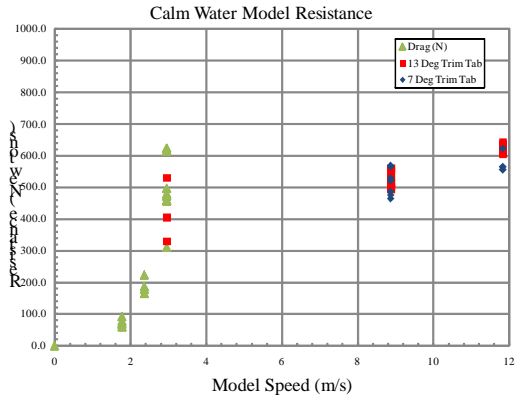

**Figure 3:** Model resistance versus speed, with and without trim tabs (7 and 13 degree tab angle), in calm water.

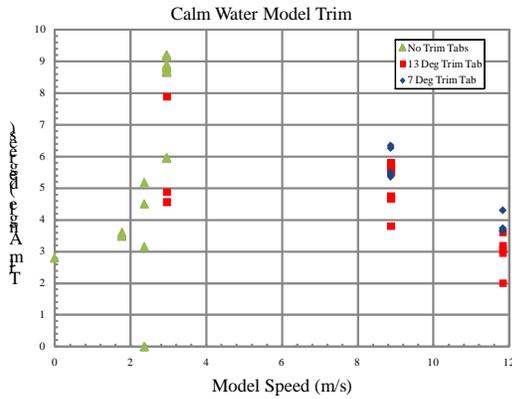

**Figure 4:** Model trim versus speed, with and without trim tabs (7 and 13 degree tab angle), in calm water.

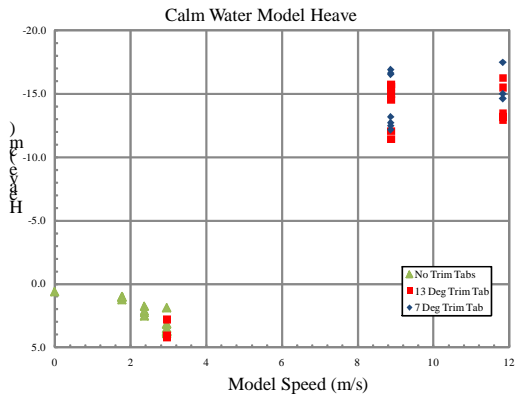

**Figure 5:** Model heave versus speed, with and without trim tabs (7 and 13 degree tab angle), in calm water. Heave is negative upward, positive downward.

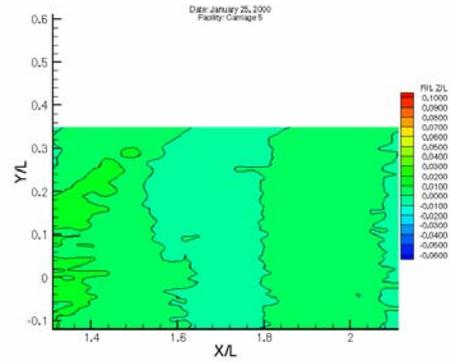

a) 1.8 m/s (5.8 ft/s)

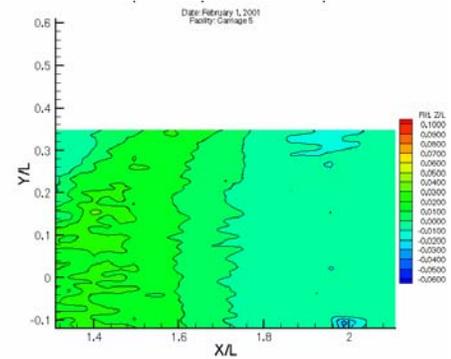

b) 2.4 m/s (7.8 ft/s)

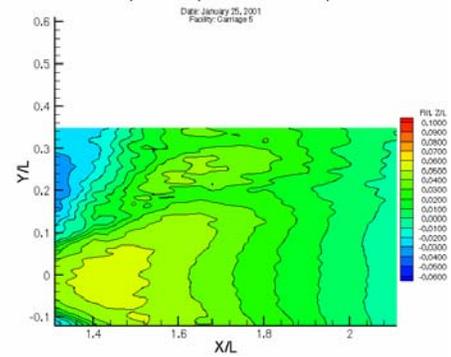

c) 3 m/s (9.7 ft/s)

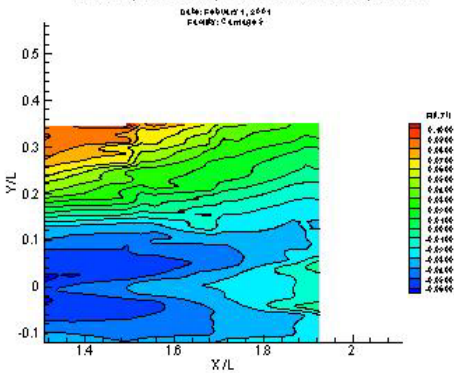

d) 9 m/s (29.2 ft/s)

**Figure 6**: Stern topologies for model speeds of 1.8, 2.4, 3, and 9 m/s (5.8, 7.8, 9.7 and 29.2 ft/s), no trim tabs.

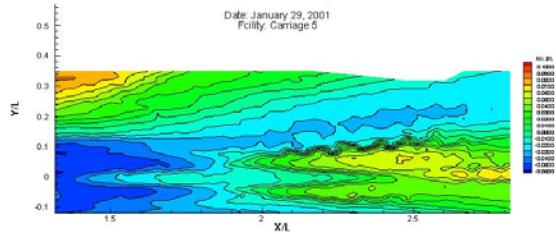

a) 7 degree tab angle

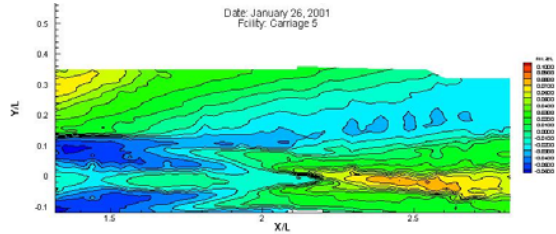

b) 13 degree tab angle

**Figure 7:** Stern wave free-surface topologies for 9 m/s (29.2 ft/s) for 7 and 13 degree trim tab angles.

**Bow Wave Topology**

The bow wave of Deep-V monohull planing boats is typically characterized by the large spray sheets, see Figure 8. Details of this feature are shown in Figures 10 and 11. Figure 9 shows the bow wave generated by the model at a speed of 3 m/s (9.7 ft/s). Note that there are certainly scale effects, less breaking and spray. Figure 10 is an image of the bow wave spray region illuminated by laser sheet, note the turbulent thin fluid sheet generated by the bow wave.

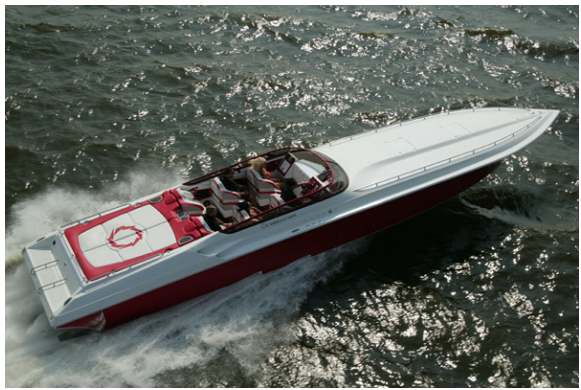

**Figure 8**: Image of a typical Deep-V monohull planing boat, showing the generated wave fields and spray.

**Table 3**: Axial Location (Station Lines) of the Laser Sheet for each Set of Conditions.

| Speed (m/s) | Axial-Location Station | Number of Runs |
|---|---|---|
| 1.8 | 3.7 | 3 |
| 1.8 | 5.5 | 2 |
| 1.8 | 11.2 | 2 |
| 1.8 | 12.35 | 2 |
| 1.8 | 12.98 | 2 |
| 2.4 | 3.7 | 1 |
| 2.4 | 5.5 | 1 |
| 2.4 | 11.2 | 1 |
| 2.4 | 12.35 | 1 |
| 2.4 | 12.98 | 1 |
| 3 | 3.7 | 3 |
| 3 | 5.5 | 2 |
| 3 | 10.93 | 1 |
| 3 | 11.2 | 1 |
| 3 | 12.35 | 1 |
| 3 | 12.98 | 1 |
| 9_7 | 6 | 3 |
| 9_7 | 6.6 | 2 |
| 9_7 | 7 | 1 |
| 9_7 | 7.5 | 2 |
| 9_7 | 8.5 | 3 |
| 9_7 | 10.04 | 2 |
| 9_7 | 11.2 | 3 |
| 9_7 | 12.35 | 3 |
| 9_7 | 12.98 | 2 |
| 9_13 | 6 | 2 |
| 9_13 | 6.6 | 2 |
| 9_13 | 7 | 2 |
| 9_13 | 7.5 | 2 |
| 9_13 | 8.5 | 3 |
| 9_13 | 10.04 | 2 |
| 9_13 | 11.2 | 3 |
| 9_13 | 12.35 | 3 |
| 9_13 | 12.98 | 2 |
| 9_No_Tabs | 7 | 2 |
| 9_No_Tabs | 10.04 | 1 |
| 9_No_Tabs | 11.19 | 2 |
| 9_No_Tabs | 12.35 | 2 |
| 9_No_Tabs | 12.98 | 1 |
| 11.8_13 | 7 | 1 |
| 11.8_13 | 10.04 | 1 |
| 11.8_13 | 11.2 | 1 |
| 11.8_13 | 12.35 | 1 |
| 11.8_13 | 12.98 | 1 |

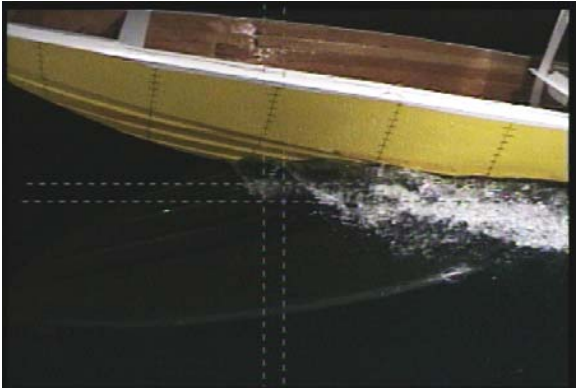

**Figure 9**: Bow wave generated by a Deep-V planing hull model at a speed of 3 m/s (9.7 ft/s).

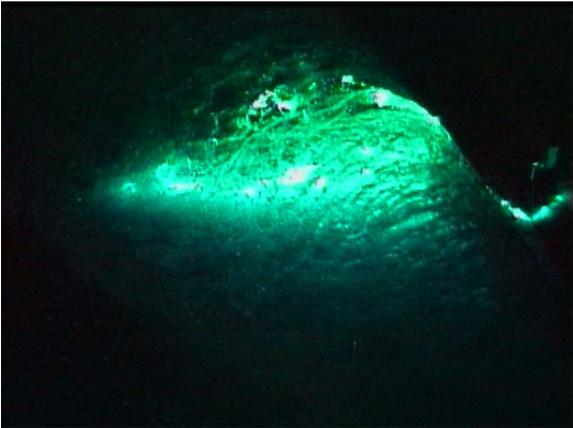

**Figure 10**: The spray region generated by the bow wave of Deep-V planing boat model illuminated by a laser sheet.

Quantitative Visualization (QVIZ), a laser sheet optical measurement technique, was utilized to measure the time-averaged transverse profiles of free surface elevation. Table 3 shows the axial locations of these measurements for the various conditions tested. Figures 11-32 show the time average profiles for 1.8, 2.4, 3, and 9 m/s (5.8, 7.8, 9.7, and 28.2 ft/s), for a number of axial locations. The RMS for each point measured is also shown for points where sufficient information was available. The regions with large RMS values are regions of large unsteadiness/breaking. Note that due to the large axial separation between profiles, no attempt was made to generate a contour map of the free-surface elevation from this data.

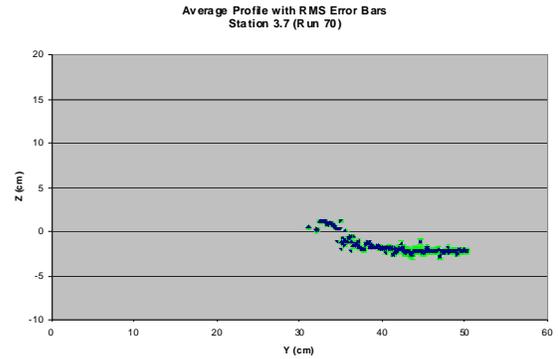

**Figure 11**: Time-averaged bow wave profile at Station 3.7 generated by a Deep-V planing craft model at a speed of 1.8 m/s (5.8 ft/s).

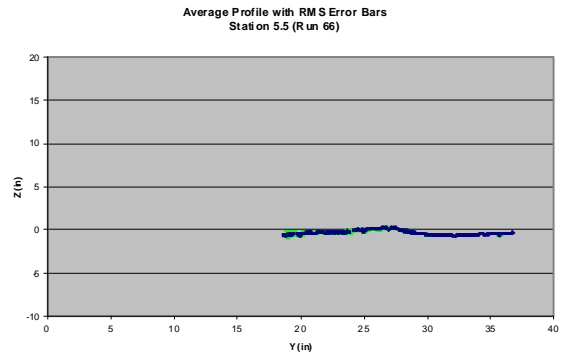

**Figure 12**: Time-averaged bow wave profile at Station 5.5 generated by a Deep-V planing craft model at a speed of 1.8 m/s (5.8 ft/s).

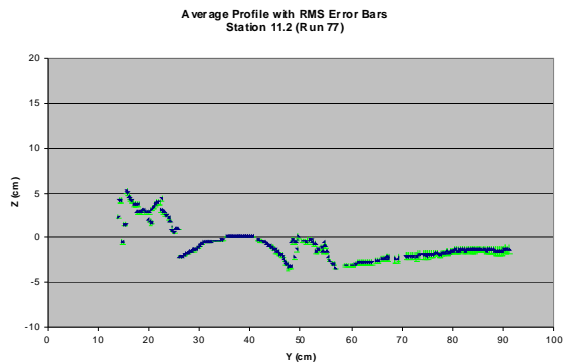

**Figure 13**: Time-averaged bow wave profile at Station 11.2 generated by a Deep-V planing craft model at a speed of 1.8 m/s (5.8 ft/s).

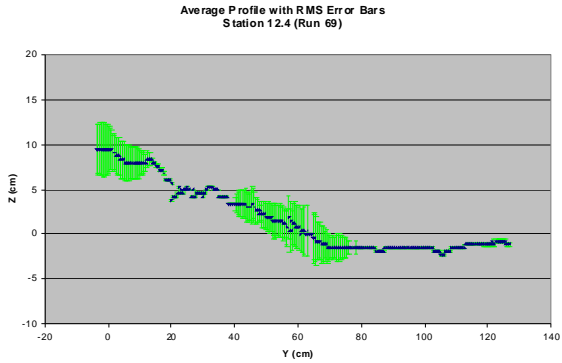

**Figure 14**: Time-averaged bow wave profile at Station 12.4 generated by a Deep-V planing craft model at a speed of 1.8 m/s (5.8 ft/s).

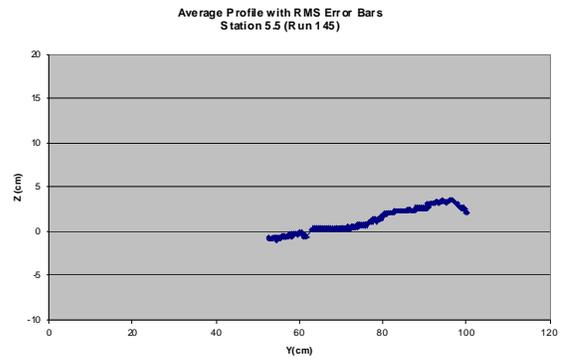

**Figure 17**: Time-averaged bow wave profile at Station 5.5 generated by a Deep-V planing craft model at a speed of 2.4 m/s (7.8 ft/s).

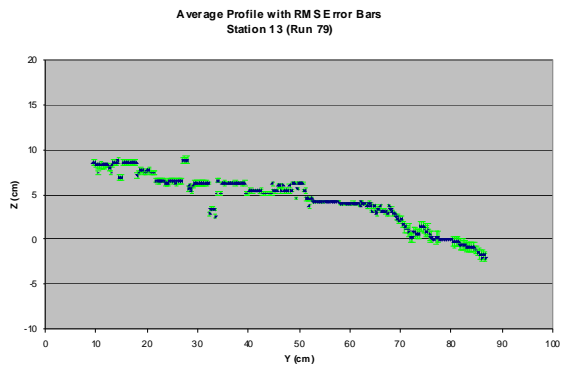

**Figure 15**: Time-averaged bow wave profile at Station 13 generated by a Deep-V planing craft model at a speed of 1.8 m/s (5.8 ft/s).

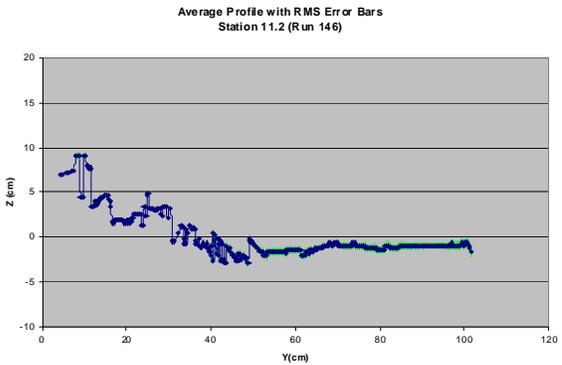

**Figure 18**: Time-averaged bow wave profile at Station 11.2 generated by a Deep-V planing craft model at a speed of 2.4 m/s (7.8 ft/s).

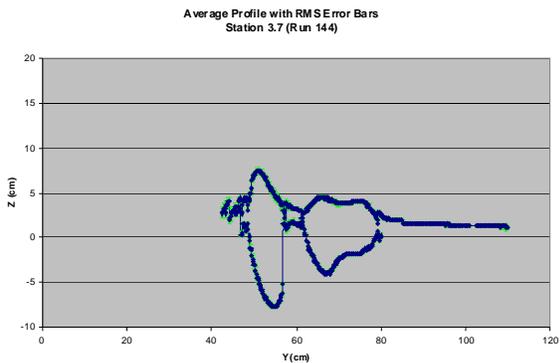

**Figure 16**: Time-averaged bow wave profile at Station 3.7 generated by a Deep-V planing craft model at a speed of 2.4 m/s (7.8 ft/s).

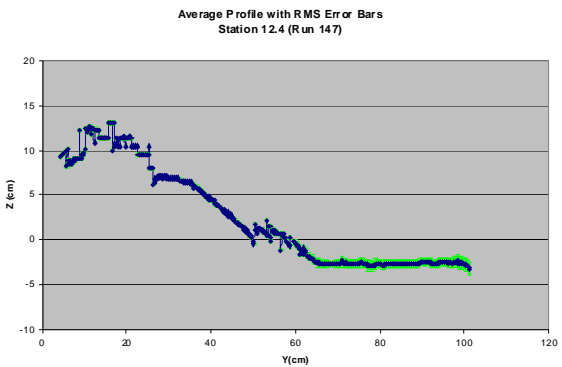

**Figure 19**: Time-averaged bow wave profile at Station 12.4 generated by a Deep-V planing craft model at a speed of 2.4 m/s (7.8 ft/s).

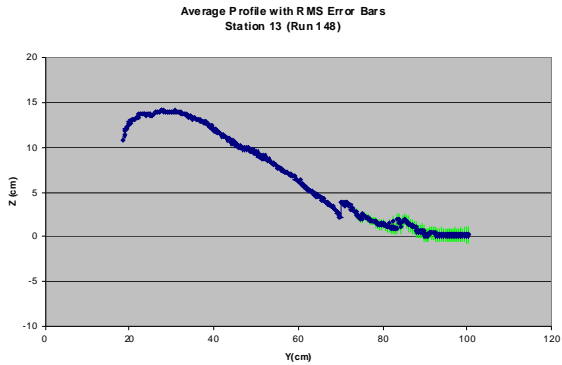

**Figure 20**: Time-averaged bow wave profile at Station 13 generated by a Deep-V planing craft model at a speed of 2.4 m/s (7.8 ft/s).

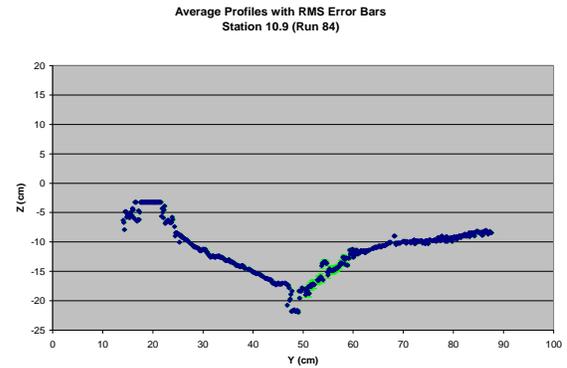

**Figure 23**: Time-averaged bow wave profile at Station 10.9 generated by a Deep-V planing craft model at a speed of 3 m/s (9.7ft/s).

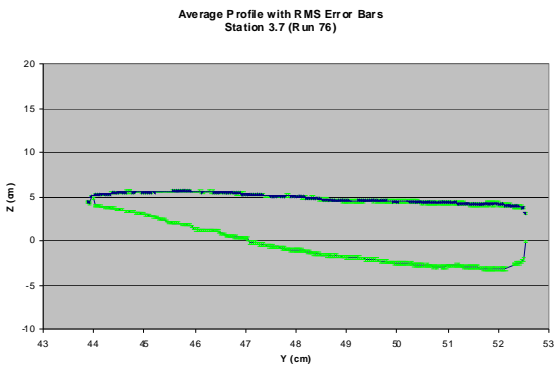

**Figure 21**: Time-averaged bow wave profile at Station 3.7 generated by a Deep-V planing craft model at a speed of 3 m/s (9.7ft/s).

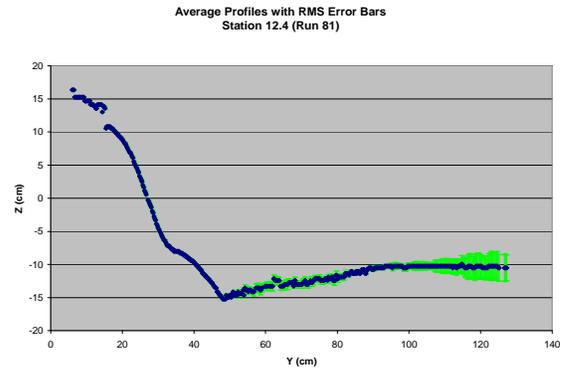

**Figure 24**: Time-averaged bow wave profile at Station 10.9 generated by a Deep-V planing craft model at a speed of 3 m/s (9.7ft/s).

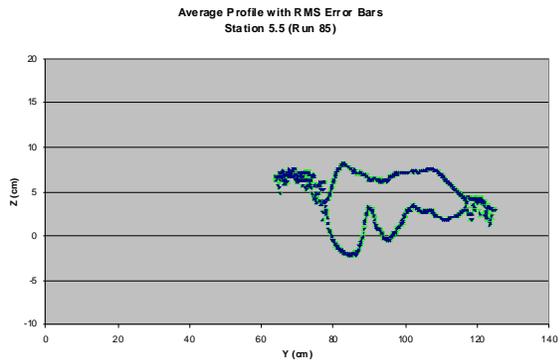

**Figure 22**: Time-averaged bow wave profile at Station 5.5 generated by a Deep-V planing craft model at a speed of 3 m/s (9.7ft/s).

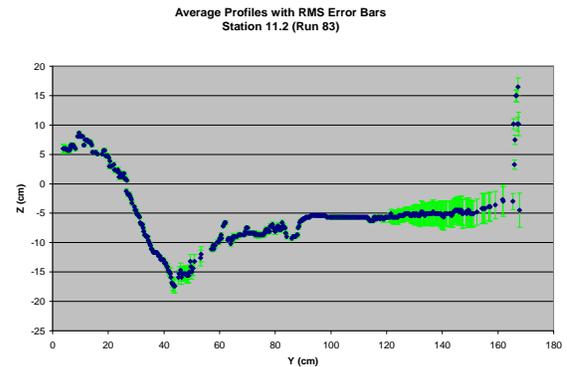

**Figure 25**: Time-averaged bow wave profile at Station 11.2 generated by a Deep-V planing craft model at a speed of 3 m/s (9.7ft/s).

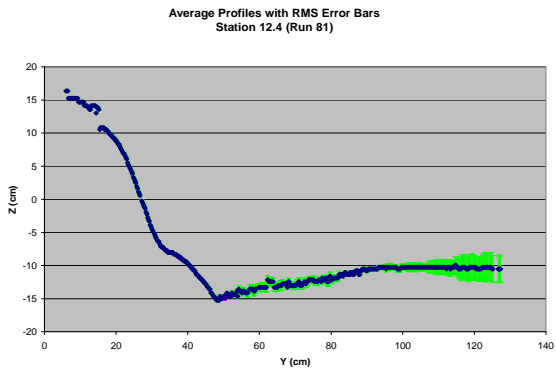

**Figure 26**: Time-averaged bow wave profile at Station 12.4 generated by a Deep-V planing craft model at a speed of 3 m/s (9.7ft/s).

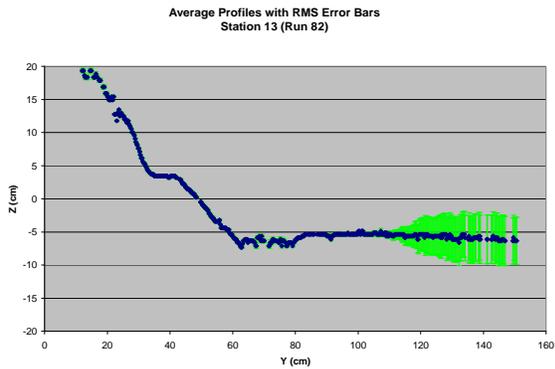

**Figure 27**: Time-averaged bow wave profile at Station 13 generated by a Deep-V planing craft model at a speed of 3 m/s (9.7ft/s).

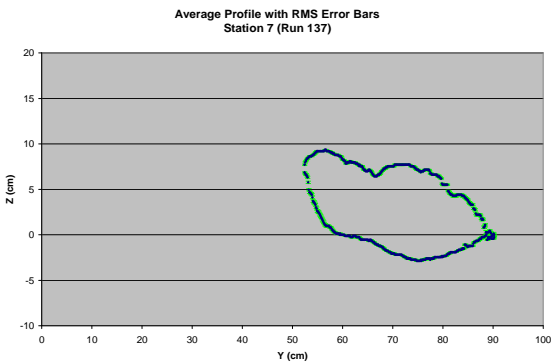

**Figure 28**: Time-averaged bow wave profile at Station 7 generated by a Deep-V planing craft model at a speed of 9 m/s (29.2ft/s), no tabs.

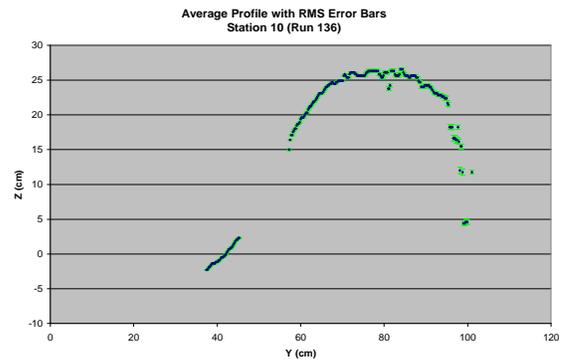

**Figure 29**: Time-averaged bow wave profile at Station 10 generated by a Deep-V planing craft model at a speed of 9 m/s (29.2ft/s), no tabs.

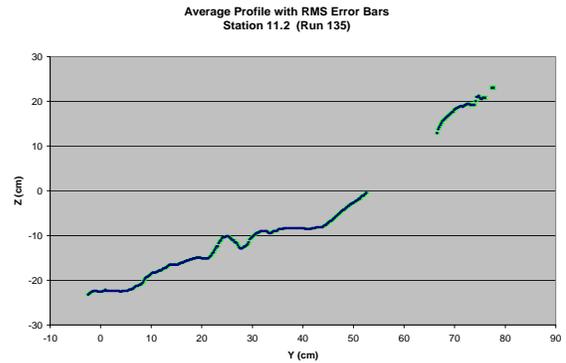

**Figure 30**: Time-averaged bow wave profile at Station 11.2 generated by a Deep-V planing craft model at a speed of 9 m/s (29.2ft/s), no tabs.

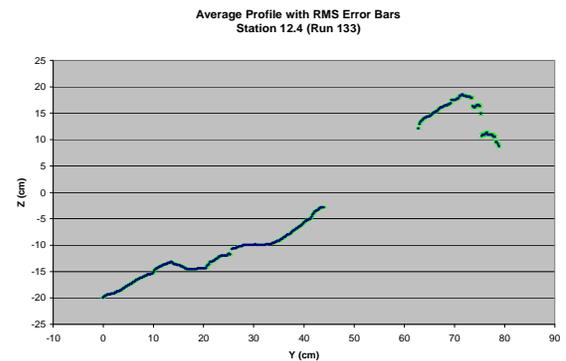

**Figure 31**: Time-averaged bow wave profile at Station 12.4 generated by a Deep-V planing craft model at a speed of 9 m/s (29.2ft/s), no tabs.

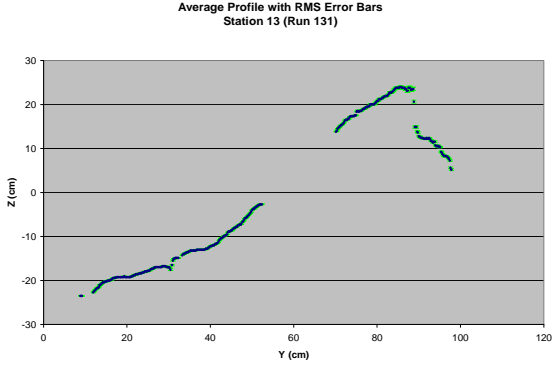

**Figure 32**: Time-averaged bow wave profile at Station 13 generated by a Deep-V planing craft model at a speed of 9 m/s (29.2ft/s), no tabs.

**NUMERICAL COMPUTATIONS**

**Formulation**
Numerical simulation of the flow field around a Deep-V planing craft was performed utilizing the Numerical Flow Analysis (NFA) code. NFA provides turnkey capabilities to model breaking waves around a ship, including both plunging and spilling breaking waves, the formation of spray, and the entrainment of air. NFA uses a cartesian grid formulation with a cut-cell representation of the hull and volume-of fluid (VOF) interface capturing of the free surface. A complete discussion of the formulation of NFA can be found in Brucker, O'Shea & Dommermuth (2010).

The free-surface boundary layer is not resolved in VOF simulations at high Reynolds numbers with large density jumps such as between air and water. A numerical breakdown is associated with the jump that occurs in the tangential velocity across the free surface. In VOF simulations, the velocity jump tends to occur right at the free-surface interface. As a result, unphysical tearing tends to occur even with high-order VOF advection schemes. In reality there is a viscous boundary layer that makes the transition from the water velocity slightly beneath the free surface to the air velocity slightly above the free surface. Smoothing and/or filtering are required to reduce the jump in the tangential velocity that occurs at the free-surface interface.

We have tested several types of smoothing filters, including a density smoother, a Smagorinsky turbulence model, a velocity smoother with projection, and a density-weighted velocity smoother with projection. We found the density weighted velocity filtering to be most effective. The complete formulation for this smoother is as follows:

$$\tilde{u}_i = \frac{\langle \rho u_i \rangle}{\langle \rho \rangle} \quad \text{for } \alpha \geq 0.5, \quad (1)$$

where $\tilde{u}_i$ is the smoothed velocity field, $u_i$ is the unfiltered velocity field, $\rho$ is the density, and $\alpha$ is the volume fraction. Brackets denote smoothing.

$$\langle F(x) \rangle = \int_{\overline{v}} W(\overline{x}) F(x - \overline{x}) dv. \quad (2)$$

Here, $F(x)$ is a general function, v is a control volume that surrounds a cell, and $W(x)$ is a weighting function. To date, we have used $W(x) = 1$ with dv over $(2\Delta x)^3$ to $(5\Delta x)^3$, where $\Delta x$ denotes cell length. $dv = (2\Delta x)^3$ corresponds to a 3-point smoother or top-hat filtering. $dv = (5\Delta x)^3$ takes out a lot of energy. Note that we require that $W$ does not either overshoot or undershoot the maximum and minimum allowable density.

Due to the high-density ratio between water and air, the preceding filtering with density weighting tends to push the water particle velocity at least one grid cell into the air. This gives us the desired physical effect that water drives air. Moreover, any velocity smoothing that does occur is limited to regions that are primarily air because we only apply the filter for $\alpha \geq 0.5$. Once the velocity is filtered, we need to project it back onto a solenoidal field in the fluid volume (V):

$$u_i = \tilde{u}_i - \frac{1}{\rho} \frac{\partial \phi}{\partial x_i} \quad \text{in V}, \quad (3)$$

where $\phi$ is a potential function. For an incompressible flow, we require

$$\frac{\partial u_i}{\partial x_i} = 0 \quad \text{in V}. \quad (4)$$

We use free-slip boundary conditions on the surface of the body (S),

$$u_i n_i = U_n \quad \text{on S}. \quad (5)$$

Here, $n_i$ is the normal to the surface of the body and $U_n$ is the prescribed normal velocity. Substituting Equation (3) into (4), yields the projection operator:

$$\frac{\partial}{\partial x_i} \frac{1}{\rho} \frac{\partial \phi}{\partial x_i} = \frac{\partial \tilde{u}_i}{\partial x_i} \quad \text{in V.} \quad (6)$$

Substituting Equation (3) into (5), gives a Neumann boundary condition on the body:

$$\frac{n_i}{\rho} \frac{\partial \phi}{\partial x_i} = \tilde{u}_i n_i - U_n \quad \text{on S.} \quad (7)$$

Since the water to air density ratio is 1000, the water velocity is weighted more heavily than the air velocity. As a result, density-weighted smoothing tends to push the water-particle velocity into the air. This provides the desirable physical effect that water particles drive the flow in the air. This smoothing method greatly reduces the presence of non-physical spray. We typically apply the filtering every 20 time steps. We have applied the preceding filtering operation to various ship-wave problems with strong shear on the free surface.

To demonstrate the effectiveness of the smoother a simple standing-wave test case was utilized. See Longuet-Higgins & Dommermuth (2001) for a complete discussion of the set-up and numerical simulations of this case. Figure 33 shows the initial conditions and cutting plane used to generate cuts shown in Figure 34. Two three-dimensional simulations have been performed with $256^3$ grid points, one with no smoothing and one with 3-point density-weighted velocity smoothing. The domain size is $1 \times 1 \times 1$, with a water depth of 0.5. The numerical simulations have been with run for 3200 times steps with a time step equal to 0.000625. The numerical simulation with no smoothing is torn up, whereas the simulation with smoothing captures the fine-scale details of the jet formation and collapse. Figure 34 shows the effects of smoothing versus no smoothing.

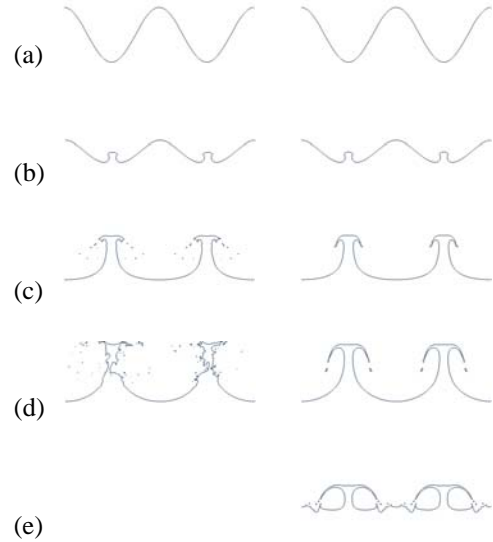

(a)

(b)

(c)

(d)

(e)

**Figure 34**: No smoothing on left versus smoothing on right. Simulation with no smoothing crashed after t=1.5. (a) t=0, (b) t=0.5, (c) t=1.0, (d) t=1.5, and (e) t=2.0.

In order to demonstrate convergence, cases of varying grid density are compared. Figure 35 shows the comparison of three different NFA simulations and a boundary integral method. The blue line indicates the boundary integral methods prediction of the free surface. The black, red and green lines indicate the 50% iso-contour of the volume fraction from NFA at $256^3$, $512^3$ and $1024^3$ domain grid points respectively. All three of these simulations utilize the 3-point density-weighted velocity smoothing with projection. The comparison to the boundary integral method improves significantly with higher grid resolution.

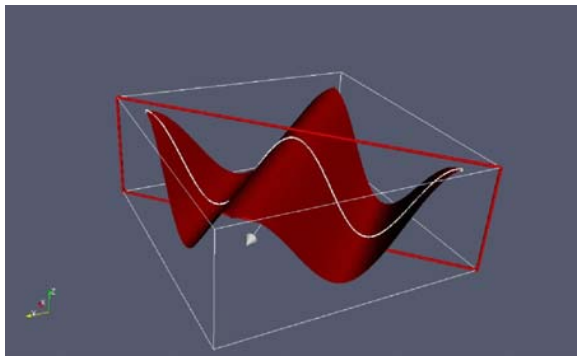

**Figure 33**: Initial conditions of standing wave test case and cutting plane used to generate cuts shown in Figure 34 outlined as red lines.

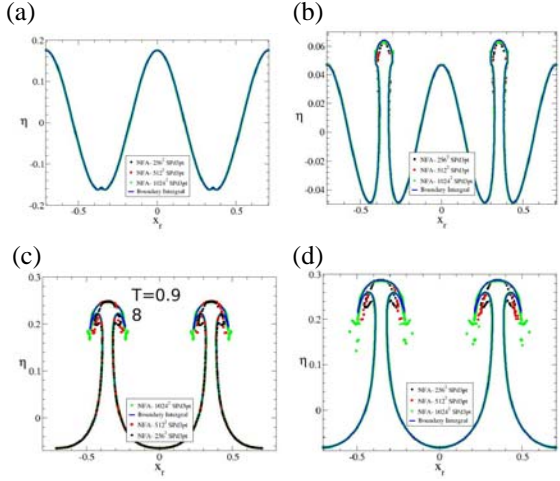

**Figure 35**: Convergence study for standing wave with velocity smoothing (a) t=0.2, (b) t=0.6, (c) t=0.9, (d) t=1.1

**Results**

Length scales are normalized by $L_o$, the length between perpendiculars. Velocity scales are normalized by $U_o$, the ship speed. As a result, time is normalized by $T_o = L_o / U_o$. Forces are normalized by $\rho_\omega U_o^2 L_o^2$, and moments are normalized by $\rho_\omega U_o^2 L_o^3$, where $\rho_\omega = 1000 kg/m^3$ is the density of water. The Froude number is $Fr = U/\sqrt{gL_o}$ where the acceleration of gravity is $g = 9.80665 m/s^2$.

Table **4** provides the normalization for the three simulations of the Deep-V model with 13-degree trim tabs at 8.91 m/s, 11.88 m/s and 14.38 m/s model scale. These speed and trim tab combinations are represented in test numbers 4 and 10 in Table 1.

**Table 4**: Normalizations.

| Experimental Test # | $L_o$ (m) | $U_o$ (m/s) | $T_o$ (s) | $Fr$ |
|---|---|---|---|---|
| 4 | 3.301 | 8.91 | 0.370 | 1.566 |
| 4 | 3.301 | 11.88 | 0.278 | 2.088 |
| 10 | 3.301 | 14.38 | 0.229 | 2.528 |

*GRIDDING*

Table 5 provides the dimensions of the computational domain for the simulations and the smallest grid spacing. For example, $X_{min}$ and $X_{max}$ respectively denote the minimum and maximum offsets along the x-axis in normalized units. Similarly, $\Delta x_{min}$ and $\Delta x_{max}$ are respectively the minimum and maximum grid spacings along the x-axis in normalized units. Note that the grid is stretched along the cartesian axes. Grid points are clustered near the ship and the mean waterline. Reflective boundary conditions are used at the tops, bottoms, sides, and fronts of the computational domains. To help waves smoothly transition out of the back of the domain, Orlanski exit boundary conditions are used downstream of the ship along the x-axis. Free-stream velocity is in the negative x direction.

**Table 5**: Deep-V Gridding Details

|     | X    | Y   | Z   | Δx      | Δy      | Δz      |
|-----|------|-----|-----|---------|---------|---------|
| Min | 0.5  | 0.0 | 0.5 | 1.2E-3  | 9.1E-4  | 1.0E-3  |
| Max | -2.5 | 1.0 | 1.0 | 1.2E-2  | 2.0E-2  | 4.0E-2  |

Table 6 provides details of how the computational domain has been distributed over the processors of the parallel computer. The multiplication of the number of subdomains in the X, Y and Z directions gives the total number of CPUs the simulation was run on, since each processor gets one subdomain. The multiplication of the number of cells per subdomain results in the total number of cells each processor works on. Finally, the number of cells in each coordinate direction multiplied together gives the total number of discrete cells in the simulation.

**Table 6**: Discretization

|       | Subdomains | Cells/Subdomains | Cells       |
|-------|------------|------------------|-------------|
| X     | 24         | 64               | 1,536       |
| Y     | 8          | 128              | 1,024       |
| Z     | 3          | 128              | 384         |
| Total | 576        | 1,048,576        | 603,979,776 |

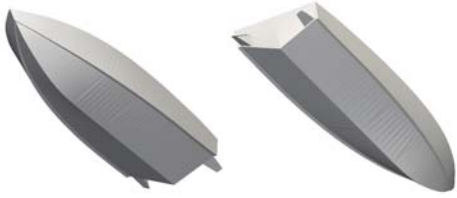

**Figure 36**: Deep-V panelization

*PANELIZATION*

Figure 36 shows the panelizations that were used for Deep-V simulations. This geometry was carefully redone to model the fine detail of the Deep-V planing hull, including all the steps, spray rails and trim tabs. The panel densities are higher in regions of high curvature. The panelization is coarse in regions where the hull geometries are flat. The panel density does not affect the accuracy of the calculation of signed-distance function that NFA uses internally to represent a ship.

Signed-distance functions are used to represent ship hulls internally within NFA. First, the shortest distance between a point in the cartesian grid to the ship hull is calculated. Then this distance is assigned a negative distance if the point is within the hull and a positive distance if the point is outside the hull to create a signed-distance function. A zero distance corresponds to a point that is on the ship hull.

The sinkage and trim of the model were deduced from the model test experiments. Only relative sinkage was recorded for each run, so the absolute sinkage was calculated from the static trim and displacement. Using 3D modeling software, the boat was oriented at the static trim and then moved vertically until the weight of the displaced volume equaled the weight of the boat. From this baseline position the relative sinkage at the bow and stern was applied. This method introduces additional errors into the sinkage and trim. Planing boats are especially sensitive to slight changes in sinkage and trim, which can greatly affect force calculation as discussed is the subsequent forces section.

*SIMULATIONS*

The numerical simulations were run on the SGI ICE on 576 CPUs for over 20,000 time steps with a non-dimensional time step Δt=0.0025. The results of the 14 m/s (46.8 ft/s), 11.8 m/s (39 ft/s) and 9 m/s (29.2 ft/s) simulations can be seen in Figures 37, 38, and 39. These figures show two isosurfaces of the time-averaged volume fraction taken over 4000 time steps or 1 boat length. The opaque blue isosurface represents a volume fraction of 0.8 while the transparent white isosurface represents a value of 0.05. Spray is highlighted in these plots since the averaged volume fraction is diffuse in the presence of highly time-varying free surface. For comparison, Figure 40 shows a photograph from the 9 m/s (29.2 ft/s) experimental test taken from approximately the same angle. Qualitative structure of the wake and spray generation agree remarkably well. We note that NFA's spray does not carry as far downstream due to grid stretching in the x and y directions. The spray can only exist as long as there are cells small enough to resolve it.

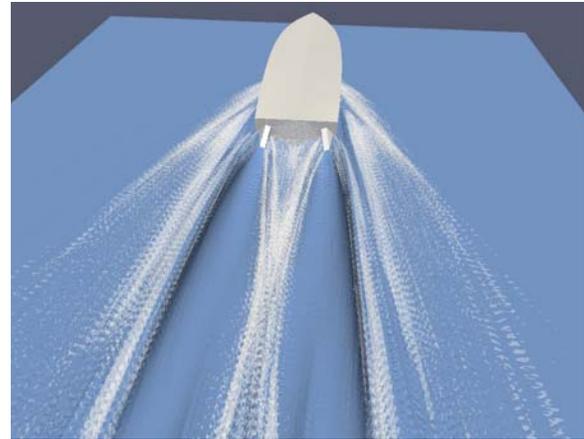

**Figure 37**: NFA simulation of Deep-V at 14 m/s (46.8 ft/s).

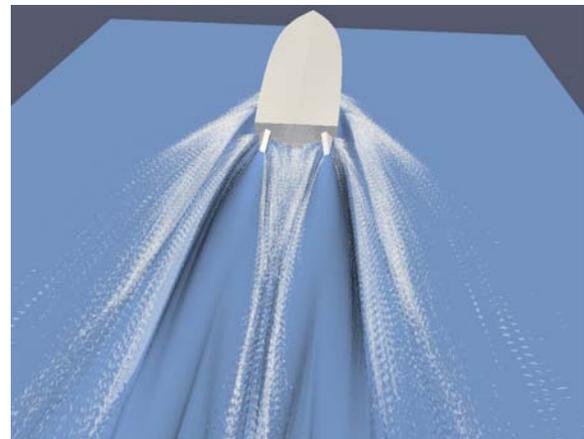

**Figure 38**: NFA simulation of Deep-V at 11.8 m/s (39 ft/s).

Whisker-probe data provides an excellent way to validate NFA's free-surface topology in the region behind the stern. Figure 41 shows the experimental data plotted on top of the NFA simulation for the 9 m/s (29.2 ft/s), 13 degree trim-tab case. The whisker-probe stern topology is outlined with a black box. The simulation compares extremely well to the experiments for such a complex flow and difficult to measure flow field.

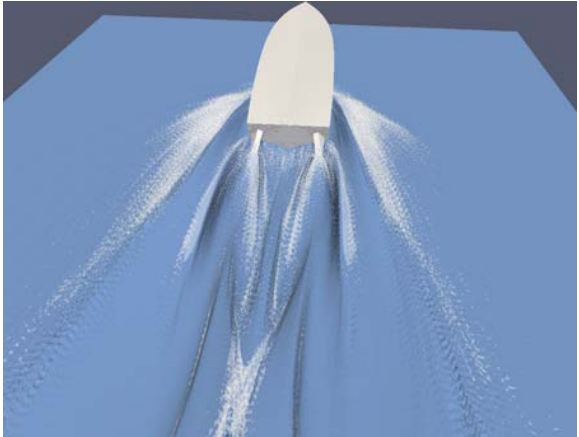

**Figure 39**: NFA simulation of Deep-V at 9 m/s (29.2 ft/s).

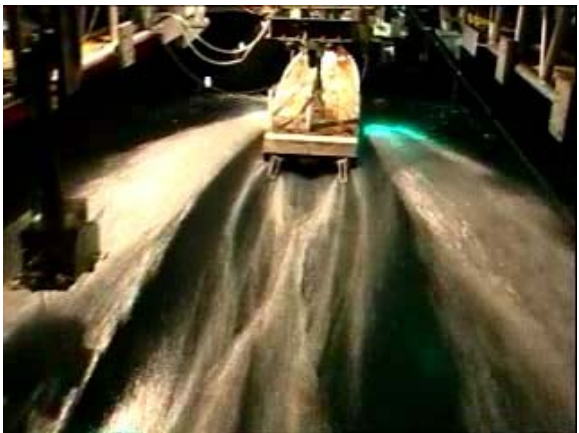

**Figure 40**: Deep-V Experimental Photograph.

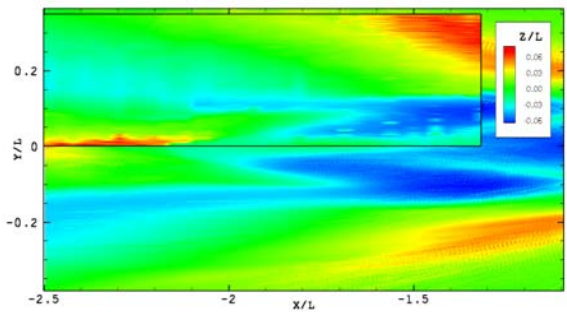

**Figure 41**: Whisker Probe and NFA comparison for 9 m/s (29.2 ft/s).

*PRESSURES*

Figure 42 shows the pressure from NFA interpolated onto the panelized representation of the hull. The bulk of the pressure acting on the hull is concentrated at the contact line as is typical in planing craft. The trim tabs also carry a large amount of pressure. The shallow angle between the tabs and water's surface means the majority of the force is orthogonal to the drag, which makes them effective at changing the trim without adding a disproportional amount of resistance. The distribution of pressure changes with speed. As speed increases the trim of the boat decreases and the dynamic lift pushes the hull out of the water.

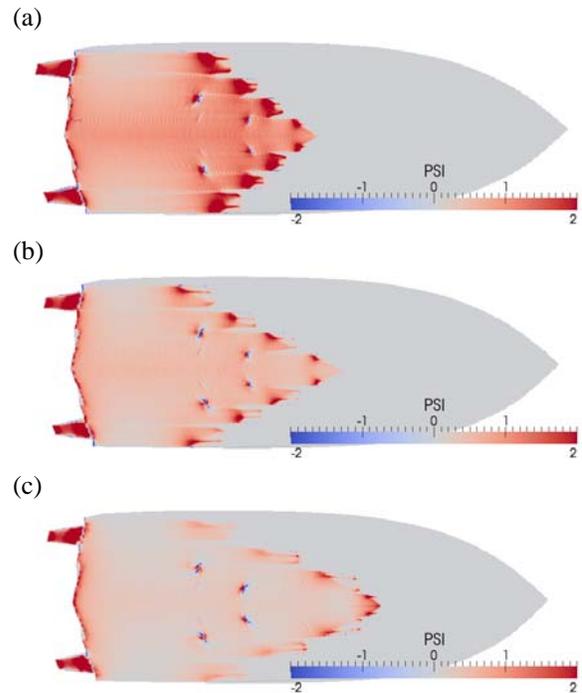

**Figure 42**: NFA prediction of pressure on the Deep-V (a) 9 m/s (29.2 ft/s), (b) 11.8 m/s (38 ft/s), (c) 14 m/s (46.8 ft/s).

*FORCES*

Forces on a model are typically broken into two parts, wave-making and viscous drag. Integrating the pressure on the hull gives us the wavemaking portion of the force. The viscous portion is calculated from the wetted surface area, Reynolds number and the ITTC flat plate friction equation (see Dommermuth, et al. 2007 and O'Shea, et al., 2008). Plots of the X and Z components of these forces are given in Figure 43. The drag measurements, represented by the red line, compare quite well with NFA's X force measurements, represented by the black line. The model experiment did not measure Z force, so the reported static displacement of the model is shown as the pink line in the figure. It should be noted that there is some uncertainty with this number. The typical behavior of a planing craft is to reduce trim and reduce sinkage as speed increases. The 14 m/s (46.8 ft/s) case actually had more sinkage than the 11.8 m/s (38 ft/s) case, which could be due to additional weight being added to help with thrust unloading as described earlier in the experimental section. The variability of the Z forces from NFA could also be attributed to the errors introduced by estimating the sinkage and trim as discussed in the previous section.

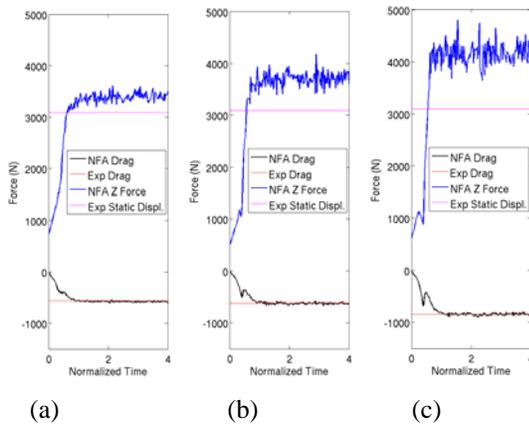

(a)          (b)          (c)

**Figure 43**: Deep-V Force Comparison (a) 9 m/s (29.2 ft/s), (b) 11.8 m/s (38 ft/s), (c) 14 m/s (46.8 ft/s).

*SPRAY*

The physics of the formation and break up of the turbulent fluid sheet associated with the generation of bow spray is a complex multiphase flow problem, acting over a large range of scales. The length scales involved vary from the small radii of spray droplets and turbulence length scales to the larger wavelengths associated with ship generated gravity waves. Sur and Chevalier (2004, 2006) measured the spray characteristic of displacement hullform bow waves, while more recently Beale et al (2010) has made measurements of the bow spray of planing craft. While visual observations seem to indicate similar formation mechanism between displacement and planing hullforms, differences are also apparent as the fluid sheets break up directly into fine spray droplets on planning craft, whereas the slower displacement hull forms appear to fragment into ligaments before spray droplet generation. Because of the physics involved and the wide range of physical scales, the numerical simulation of the spray generated by a high-speed planing vessel is a complex and challenging problem. Visualizing the results of a simulation involving large amounts of spray can also be difficult. Figure 45 represents one way of looking at NFA's prediction of spray. The plots are of transverse cuts taken at varying distances from the transom. Three speeds and three longitudinal locations are shown. From top to bottom the speeds are 9 m/s (29.2 ft/s), 11.8 m/s (39 ft/s), and 14 m/s (46.8 ft/s), while the longitudinal locations from left to right are 0.2 ship lengths forward of the transom, at the transom and 0.2 ship lengths aft of the transom. The plots show the averaged volume fraction over 4000 time steps, or 1 boat length. The blue represents a volume fraction of 0.0, or 100% air and the red represents a volume fraction of 1.0, or 100% water. An instantaneous snapshot of the volume fraction would show a discrete jump from air to water. Alternatively, when averaged through time, the volume fraction becomes diffuse in regions of high variability. The white region between the red and blue represents any volume fraction between 0.0 and 1.0 and serves to envelope the region in which spray exists. Figure 44 represents another way of looking at the same time averaged data set. Two volume fraction isosurfaces are displayed. The opaque blue isosurface represents a value of 0.8 while the transparent white isosurface represents a value of 0.05. The white surface serves as an upper bound for the spray envelope. At this viewing angle the width and depth of the spray cloud becomes visible and is qualitatively similar to experimental observations.

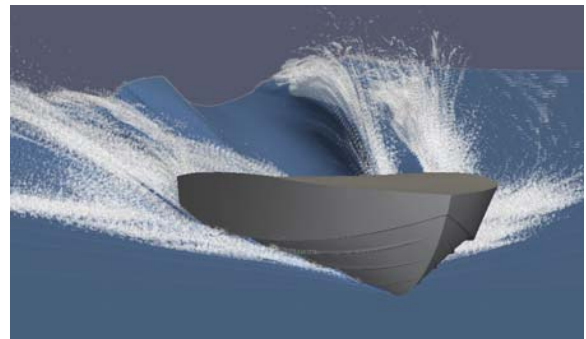

**Figure 44**: Deep-V Front View

## CONCLUSIONS

The model test of a Deep-V, monohull model provided a comprehensive set of planing craft data for the validation of CFD codes. The test generated several unique types of data, including traditional resistance and trim measurements, hull pressures, and multiple quantifications of the flow field itself. Each of these data types represented a specific interpretation of planing craft hydrodynamics, and collectively, this information should prove valuable as CFD codes are developed and evaluated for planing craft applications.

With regard to conclusions specific to the model test itself, there were several conclusions to note. First, there was enormous discrepancy between the estimated quantity of unloading weight and moment correction weight required. It is not understood why initial moment correction methods were unsuccessful at producing realistic trimming characteristics of the model. Further study of the influences affecting trim angle of planing models is necessary.

Hull pressure measurements proved very difficult to obtain. Recorded data did not appear to be valid, and highlights the need for more study in this area. A probable cause of the problem is poor installation and/or calibration of the gages. The ability to obtain accurate pressure readings in a planing boat model test would be a valuable capability, both for the assessment and validation of CFD codes and for the development of future advanced marine vehicles.

Although it was not possible to perform traditional wave pattern resistance analysis on the longitudinal wave cut data, due the presence of wave reflection, the wave-cut records provide a valuable data set for comparison with CFD code predictions.

Whisker-probe technology was applied in the extremely challenging task of mapping the stern wake of a Deep-V planing hull. The variation in characteristics of this stern wave throughout the required speed range required significant modification to the existing whisker-probe system, both in the addition of whisker probes, and the incorporation of additional travel in the vertical direction. These modifications were successfully integrated, stern topography measurements proceeded without incident, and detailed flow-field surface maps were produced. It is anticipated that these data products will prove extremely useful in the validation of similar maps output from CFD codes.

Time-averaged calculations quantifying the envelope of the bow-spray region were accomplished using a Quantitative Visualization method. This initial attempt at quantifying the flow field of planing craft highlights the necessity for technique refinement prior to future quantification of planing craft bow-wave spray data. These tests showed that, in order to achieve a steady average image, a larger image sample set is required as the flow becomes more unsteady. Additionally, the capability to perform unsteady analysis of this type of data is needed.

The difficulty in modeling planing craft using CFD necessitated the exploration of various smoothing and filtering methods to handle spurious spray generation. The smoothing procedure developed for this work is invaluable in modeling high speed planing craft, and any boat that generates a large amount of spray. An even more robust and physics based solution is being developed and would resemble the wall boundary layer model that is reported in Rottman, et al., (2010).

## ACKNOWLEDGEMENTS

The Office of Naval Research supports this research. Dr. Steve Russell supported the NFA research and Dr. L. Patrick Purtell the analysis and reporting of the testing effort. SAIC IR&D also supported the development of NFA. This work is also supported in part by a grant of computer time from the Dept. of Defense High Performance Computing Modernization Program, http://www.hpcmo.hpc.mil/. The numerical simulations were performed on the SGI ICE at the U.S. Army Engineering Research and Development Center (ERDC). The authors would also like to acknowledge the contributions of those associated with the model test program: James R. Rice, Donnie Walker, Deborah Furey, Bill Boston & Peter Congedo

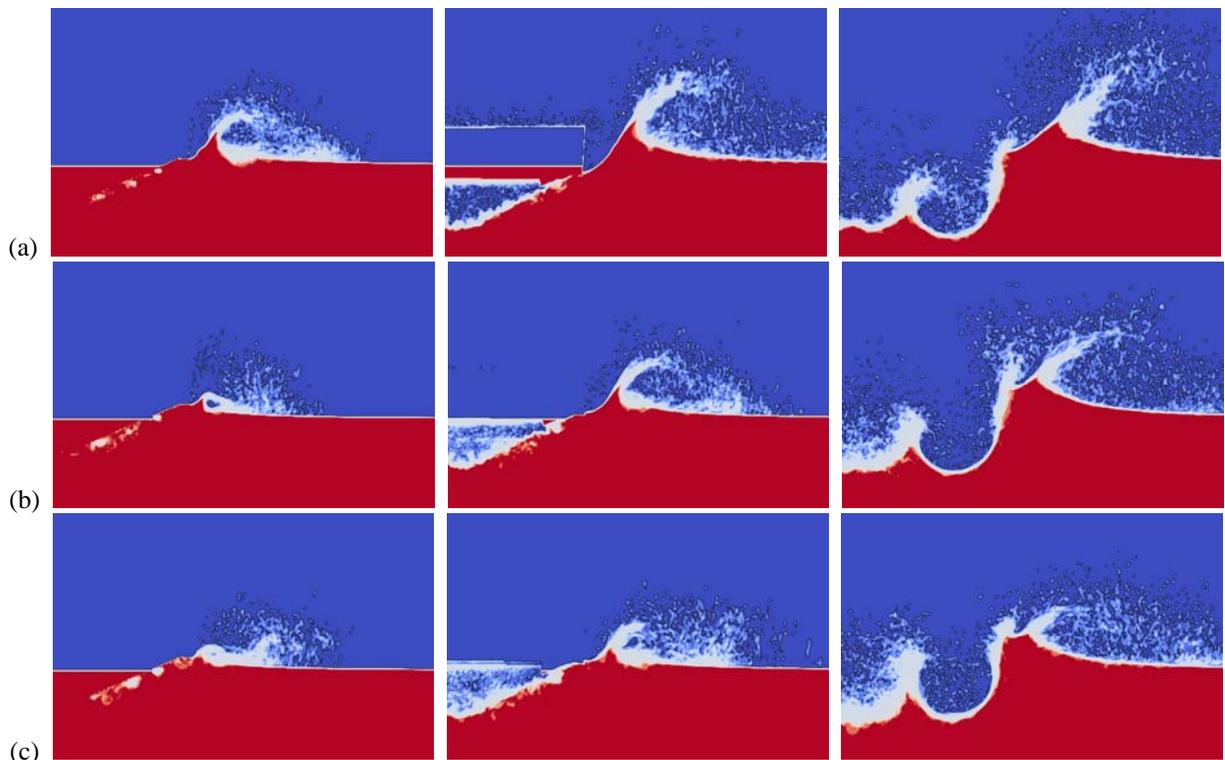

**Figure 45**: Transverse slices of Deep-V simulations (a) 9 m/s (29.2 ft/s) , (b) 11/ m/s (38 ft/s), (c) 14 m/s (46.8 ft/s). Locations at 0.2 ship lengths forward of the transom, at the transom and 0.2 ship lengths aft of the transom are shown from left and right.